\documentclass[
reprint,
notitlepage,
amsmath,
amssymb,
aps,
prb
]{revtex4-1}

\usepackage{graphicx}
\usepackage{physics}

\newcommand{\rmi}{\mathrm{i}}
\newcommand{\rme}{\mathrm{e}}

\newcommand{\mat}[1]{\boldsymbol{\mathit{#1}}}

\begin{document}

\title{Near-field-assisted capacity of spoof-plasmonic channels}

\author{Mikhail Erementchouk}%
 \email{merement@gmail.com}
 \author{Pinaki Mazumder}%
 \email{pinakimazum@gmail.com}
\affiliation{%
 Department of Electrical Engineering and Computer Science,\ 
 University of Michigan, Ann Arbor, MI 48109 USA 
}%

\begin{abstract}
  Establishing universal features of spoof-plasmonic systems beyond
  spectral properties is challenging due to the complexity of the specific
  physical realizations of spoof-plasmonic channels. We introduce a simple
  1D scalar model reproducing the key properties of spoof-plasmonic
  channels and investigate manifestations of plasmonic-like features when
  only a few local resonances are present. We show that the channel between
  the source applied to the interior of the structure and the terminal ends
  (output ports) effectively comprises two subchannels. The activation of
  one of the subchannels depends on the spatial variation of the source,
  and therefore, the contribution of this subchannel in conventional
  systems is small if the source occupies a subwavelength region. We show
  that, in spoof-plasmonic structures, the activation of this subchannel
  can enhance significantly in the frequency region where the
  spoof-plasmonic effects are prominent. This demonstrates that even a few
  local scattering resonances may strongly impact the flow of wave-carried
  information.
\end{abstract}

\maketitle

\section{Introduction}
\label{sec:intro}

A similarity between the properties of plasmon polaritons and the states of
the electromagnetic field near corrugated conducting surfaces has led to
the emergence of spoof-plasmonics~\cite{huidobroSpoof2018, tangConcept2019,
  garcia-vidalSpoof2022}. Propagation of waves in spoof-plasmonic
structures mimics plasmon polaritons opening ways for achieving plasmonic
effects across broad frequency intervals, including those that are far
below the plasmon frequencies of conventional materials. The physical
origin of the emergence of plasmonic features are the resonances at open
cavities formed by the corrugations. Since the physical origin of such
resonances is secondary, spoof-plasmonics, while established originally for
electrodynamic structures~\cite{pendryMimicking2004,
  garcia-vidalSurfaces2005}, emerges as a manifestations of fundamental
properties of wave systems with local resonances. It is, therefore, not
surprising that the spoof-plasmonic approach found applications in other
wave systems besides electromagnetic, predominately
acoustic~\cite{christensenCollimation2007, christensenEnhanced2010,
  maAcoustic2016, cselyuszkaSpoofFluidSpoof2019, jankovicAcoustic2021,
  parkUltraslow2021, phamHow2023}. Moreover, the physical origin of the
local resonances may be different from the nature of propagating waves as
was explored in acoustic waveguides in Ref.~\onlinecite{parkUltraslow2021}
and in cascade quantum lasers in Ref.~\onlinecite{sunSpoof2013}, where the
resonances are formed by excitons in multiple quantum
wells~\cite{deychPolariton2000, ivcenkoOptical2005}.

At the same time, revealing universal features of spoof-plasmonic channels
is hindered by the difficulties associated with the first-principle
representation of particular physical implementations. The
common approach to deal with these difficulties is to employ a version of
the effective medium approximation in conjunction with full-wave
simulation. One of the drawbacks of the effective medium approach is hiding
the effect of local resonances. This excludes the effect of the strong
subwavelength scale spatial modulations of the field in the channel, which
may play the crucial role in particular applications~\cite{joySpoof2017,
  aghadjaniForce2018, shugayevGiant2021}, and creates an impression that
the emergence of spoof-plasmonic effects require many local resonances,
similarly to developing photonic crystal effects in modulated dielectric
structures.

To address these difficulties and to investigate the emergence of the
spoof-plasmonic effects, we introduce a simple 1D scalar model of a
spoof-plasmonic channel. The model reveals that such channels demonstrate a
plethora of effects associated with the sensitivity of spoof plasmons to
subwavelength features at frequencies near the edge of the fundamental
spoof plasmonic band. Importantly, the emergence of these effects do not
require the presence of many local resonances. In the present paper, we
focus to the effect of enhancement of a the signal induced by a source with
the spatial variation at the subwavelength scale.

The rest of the paper is organized as follows. In Section~\ref{sec:model},
we introduce the model and solve the main problems regarding the transport
properties and the system response. In Section~\ref{sec:info-cap}, we apply
these results to demonstrate the enhanced sensitivity with respect to
subwavelength variations of the external excitation.

\section{Scalar model with distributed local resonances}
\label{sec:model}

As the 1D scalar model of spoof-plasmonic channels we consider a string
with attached harmonic oscillators. Denoting the deviations of the string
by $\psi(x)$ and of the $n$-th harmonic oscillator by $\phi^{(n)}(t)$, the
equations of motion governing the spatial distribution of deviations from
equilibrium in the steady state regime,
$\psi(x, t), \phi^{(n)}(t) \propto \exp(- \rmi \omega t) $, can be written as
\begin{equation}\label{eq:string_eigen}
\begin{split}
 k^2 \psi & = - \psi'' + \eta \sum_n \delta(x - x^{(n)}) \left[ \psi(x) - \phi^{(n)} \right], \\
 \omega^2 \phi^{(n)} & = - \Omega_c^2 \left[ \psi(x^{(n)})  - \phi^{(n)} \right],
\end{split}
\end{equation}
where $k = \omega/v$ with $v$ being the speed of waves
in the string without oscillators, $\eta$ is the coupling parameter having the
wavenumber dimension, and $\Omega_c$ is the natural frequency (with which we
associate the respective wavenumber $k_c = \Omega_c/v$) of
oscillators attached to an immobilized string.

We adopt the transfer matrix approach, which we briefly overview in
Appendix~\ref{sec:tm-approach}. The consideration noticeably simplifies if
two bases are kept in mind: the Cauchy basis with the state vector
$\mat{\Psi}^{(C)}(x)$ [defined in Eq.~\eqref{eq:tm-tm_cauchy}], and the basis
based on linearly independent solutions $\mat{\Psi}^{(w)}(x)$
[Eq.~\eqref{eq:tm-funct_psi}]. We adopt the notations with the upper index
indicating the chosen basis, when the result depends on the explicit form
of the transfer matrix and the state vector, and the omitted upper index
denoting the basis-independent form.

The transfer matrices are found as the product of transfer matrices across
the intervals between the resonances and the transfer matrices across the
resonances. According to Eq.~\eqref{eq:relation_field_functions}, for
$x^{(n-1)} < x_1, x_2, < x^{(n)}$, the transfer matrix is
\begin{equation}\label{eq:free_transfer_cauchy}
 \mathcal{T}^{(C)}(x_2, x_1) = \widehat{W}_w(x_2) \widehat{W}_w(x_1)^{-1}, 
\end{equation}
where we have taken into account the transfer matrix in basis of linearly
independent solutions is identity, and $\widehat{W}_w(x)$ is the
Wronsky matrix given by Eq.~\eqref{eq:tm-Wronsky-matrix-solutions}.

The transfer matrix across the resonance is the easiest to derive in the
Cauchy basis. 
Enforcing the continuity of the string's displacement at $x = x^{(n)}$, and
expressing $\mat{\Psi}^{(C)}\left( x^{(n)} + 0 \right)$ in terms of
$\mat{\Psi}^{(C)}\left( x^{(n)} - 0 \right)$, we find the transfer matrix through
the $n$-th resonance 
\begin{equation}\label{eq:mech_x_n_cauchy_tm}
 \mathcal{T}^{(C)} \left( x^{(n)} \right) 
  = \mathcal{I} + \gamma(\omega) \mqty(0 & 0 \\ 1 & 0),
\end{equation}
where $\mathcal{I} = \widehat{1}_2$, a $2\times2$ identity matrix,
$\gamma(\omega) = \eta \omega^2/(\omega^2 - \Omega^2_c)$.
The form of the transfer matrix in the basis of independent solutions is
obtained using \eqref{eq:relation_field_functions}:
\begin{equation}\label{eq:mech_x_n_sols_tm}
 \mathcal{T}^{(w)} \left( x^{(n)} \right) 
 = \mathcal{I} + \frac{\rmi \gamma(\omega)}{2k} \mqty(-1 & - \rme^{- 2 \rmi k x^{(n)}} \\
 \rme^{2 \rmi k x^{(n)}} & 1) .
\end{equation}
We take a note of a useful basis-independent property
$\mathcal{T}\left( x^{(n)} \right)^N = \eval{\mathcal{T}\left( x^{(n)} \right)}_{\eta \to N \eta}$.

\subsection{Dispersion relation}

Most straightforwardly, the spoof-plasmonic features of the introduced
model are demonstrated by the dispersion relation governing the
excitations. To this end, we consider a periodic system with period $d$
obtained by replicating the elementary cell occupying the interval
$\left[ x_0 , x_0 + d \right]$ with the resonance located at $x^{(0)}$
strictly inside the interval. Consequently, the transfer matrix across the
period is
\begin{equation}\label{eq:mech_period_long_tm}
  \mathcal{T}^{(C)}_P  =
  \mathcal{T}^{(C)}\left( x_0 + d, x^{(0)} \right)  \mathcal{T}^{(C)}\left( x^{(0)} \right)
  \mathcal{T}^{(C)}\left( x^{(0)}, x_0 \right).
\end{equation}

By virtue of the Bloch theorem, the eigenstates of the system must satisfy
$\mathcal{T}^{(C)}_P \mat{\Psi}^{(C)}(x) = e^{\rmi \beta d} \mat{\Psi}^{(C)}(x)$, where
$\beta$ is the Bloch wavenumber. The relation between $\omega$ and $\beta$ is found as
$\det(\mathcal{T}^{(C)}_P - p) = 0$, where $ p = e^{\rmi \beta d}$, which yields the
equation
 $p^2 - p \Tr \left( \mathcal{T}^{(C)}_P \right) + \det \left( \mathcal{T}^{(C)}_P \right)  = 0$.
Taking into account that $\det \left( \mathcal{T}^{(C)}_P \right)  = 1$, we obtain
\begin{equation}\label{eq:mech_disp_long_tm}
  \cos(\beta d) = \frac{1}{2} \Tr \left( \mathcal{T}^{(C)}_P \right)  = B(\omega; d),
\end{equation}
where
\begin{equation}\label{eq:mech_disp_B_fun}
  B(\omega; d) = \cos(k d ) + \frac{\gamma(\omega)}{2k} \sin(k d).
\end{equation}

It is edifying to consider an alternative derivation of the dispersion
equation mimicking that employed while considering electrodynamic
structures~\cite{garcia-vidalSpoof2022, songActive2009,
  erementchoukElectrodynamics2016}. This approach relies on constructing
the solution to the governing equations of motion based on the Bloch
theorem prescribing
$\psi(x) = \rme^{\rmi \beta x} u(x)$, where $u(x)$ is a periodic function with
period $d$. Representing $u(x)$ as a Fourier series, we
obtain
\begin{equation}\label{eq:mech_Bloch_series}
 \psi(x) = \sum_m e^{\rmi \beta_m x} \psi_m,
\end{equation}
where $\beta_m = \beta + 2\pi m/d$ and $\psi_m$ is the amplitude of the
$m$-th Bloch component. Using this representation in
Eq.~\eqref{eq:string_eigen}, we find
\begin{equation}\label{eq:mech_system_tra_tm_0}
 \left( k^2 - \beta_m^2 \right) \psi_m = 
 	\frac{\gamma(\omega)}{d}
 				\sum_{m'} e^{\rmi 2 \pi (m' - m) x^{(n)}/d} \psi_{m'}.
\end{equation}
This yields the dispersion equation in the form
\begin{equation}\label{eq:mech_disp_tra_tm}
 1 = \frac{\gamma(\omega)}{d} \sum_{m} \frac{1}{k^2 - \beta_m^2}.
\end{equation}
While seemingly different, this dispersion equation is equivalent to the
one obtained within the transfer matrix approach
[Eq.~\eqref{eq:mech_disp_long_tm}], as we show in
Appendix~\ref{sec:app-equivalence}.

The dispersion equation written in form~\eqref{eq:mech_disp_tra_tm} has the
typical structure of dispersion equations describing the spoof plasmonic
channels~\cite{garcia-vidalSurfaces2005, songActive2009,
  erementchoukElectrodynamics2016}. It is not, therefore, surprising that
the main features of the dispersion relation governing the excitations of
the system under consideration reproduce those of other spoof-plasmonic
systems. It is instructive to start with the
single mode approximation of Eq.\eqref{eq:mech_disp_tra_tm} (the
approximation often used in studying SSPP). It corresponds to keeping only
the term $m=0$ in the series representation of the dispersion equation,
which yields
\begin{equation}\label{eq:mech-sma}
  \omega^2(\beta) = \Omega^2_c - \left(\Omega_c^2 + \eta v^2/d  \right)^2\frac{1}{2 \beta^2v^2}.
\end{equation}
This dispersion relation is similar to the surface plasmon polariton
dispersion law with the plasma frequency $\omega_p = \sqrt{2}\Omega_c$. This reveals
the main feature of the spoof plasmonic channels emerging at frequencies
close to the edge of the fundamental band: the ability to support
subwavelength features, with the smallest characteristic spatial scale
determined by the period of the structure.

At the same time, Eq.~\eqref{eq:mech-sma} demonstrates the typical drawback
of the single-mode approximation --- the incorrect prediction for the edge of
the fundamental band coinciding with the resonance
frequency~\cite{erementchoukElectrodynamics2016, schnitzerSpoof2017}. To
obtain the correct value, one needs to consider the exact dispersion
equation. In the short-period case, when at the boundary of the Brillouin
zone, take $\cos(kd) \approx 1$, $\sin(kd) \approx kd$, we find that the boundary of
the fundamental edge is shifted down from $\Omega_c$, so that the resonance
frequency is inside the fundamental gap
\begin{equation}\label{eq:mech-band-edge}
  \omega_B^2 = \frac{\Omega_c^2}{1 + \eta d/2}.
\end{equation}

Using the same approximation, it is straightforward to obtain the dispersion
curve for the whole fundamental spoof-plasmonic band:
\begin{equation}\label{eq:mech-edge-appr}
  \omega^2(\beta) = \frac{\Omega_c^2}{\sin^2(\beta d/2) + \eta d/2}.
\end{equation}

The condition determining the upper edge of the first gap, separating the
fundamental band from the higher bands, is more complex. However, if it is
far below the frequency of the free oscillations at the boundary of the
Brillouin zone, $v \beta_B$, one can employ the short period approximation
with the slight modification, $\cos(kd) \approx 1 - (kd)^2/2$, and find that the
first gap terminates at
  $\omega_g^2 = \Omega_c^2 + \eta v^2/d$.

\begin{figure}[tb]
  \centering
  \includegraphics[width=2.6in]{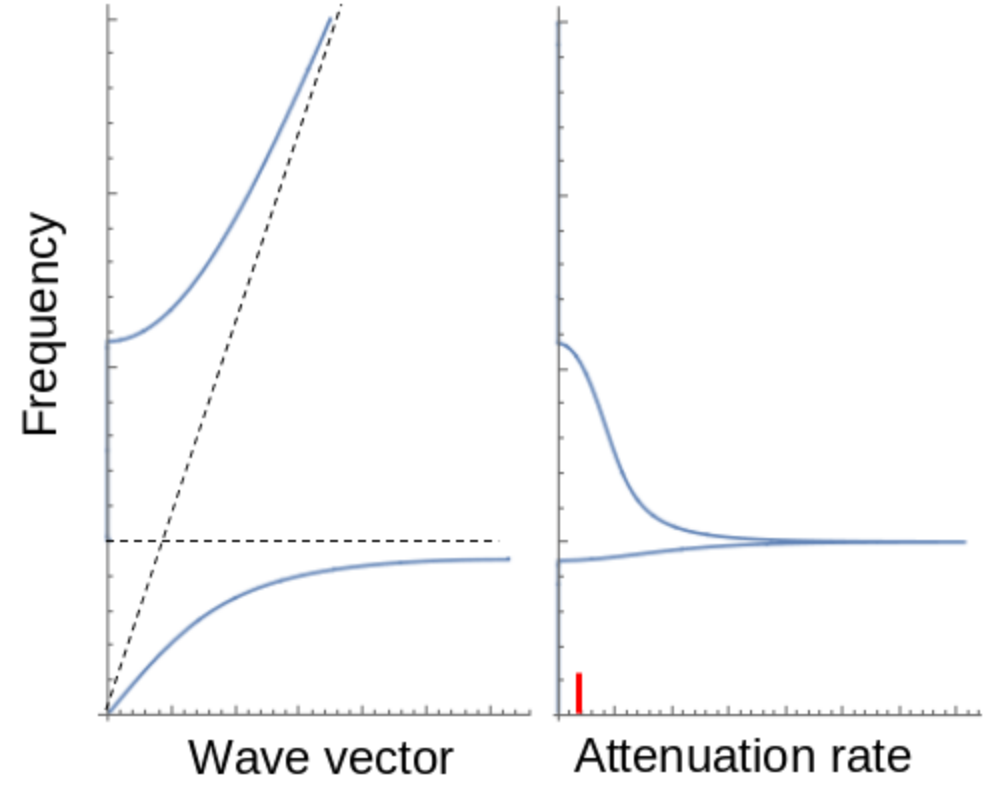}
  \caption{(Left panel) The overall view of the dispersion diagram 
    inside the first Brillouin zone. (Right panel) The attenuation along
    the structure inside the spoof-plasmonic gap. The short red line shows
    the maximum attenuation inside the gap at the boundary of the Brillouin
    zone (``the photonic crystal'' effect). Because of the chosen short
    period, the frequency corresponding to the anticrossing at the boundary of
    the Brillouin zone is the order of magnitude higher than the resonance
    frequency and, therefore, is not shown.}
  \label{fig:mech-disp-diag}
\end{figure}

The overall shape of the dispersion curves for the first 
spoof-plasmonic bands is shown in Fig.~\ref{fig:mech-disp-diag}. The
curious consequence of the resonance frequency being inside the gap is that
$\Omega_c$ splits the gap into two parts with noticeably different properties as
is evident from the spectrum of the longitudinal attenuation length (the
right panel of Fig.~\ref{fig:mech-disp-diag}). The same effect takes place
in electrodynamic SSPP as was found in~\cite{joySpoof2017} in conjunction
with the resonance tunneling effect and the dependence of the transparency
window on the geometry of the defect cell.

\subsection{The effect of decay}

In the present paper, we focus on the properties of an ideal system
without decay. However, in the realistic systems, the effect of decay
should be taken into consideration as it may significantly impact the
manifestation of spoof-plasmonic features.

To simplify the analysis, we assume that the main source of decay is the
dynamics of resonances, so that the second equation in
Eqs.~\eqref{eq:string_eigen} has the form
\begin{equation}\label{eq:mech_x_n_long_tm_damp}
  \left( \omega^2 - \Omega_c^2 + \rmi \alpha_d \omega \right) \phi^{(n)} = - \Omega_c^2 \psi(x^{(n)}).
\end{equation}

Following the derivation of the transfer matrix above, it is easy to see
that the effect of decay reduces to modifying the expression for $\gamma(\omega)$ defining the
transfer matrix through the resonance:
\begin{equation}\label{eq:damped_gamma}
  \gamma(\omega) = \eta \omega \frac{\omega + \rmi \alpha_d }{\omega^2 - \Omega_c^2 + \rmi \alpha_d \omega}.
\end{equation}
The decay leads to attenuation of spoof-plasmonic excitations along the
structure. Assuming weak decay, we find the attenuation length at the
frequency corresponding to the edge of the fundamental band
\begin{equation}\label{eq:att_length}
  l'' \approx \frac{d^2 \Omega_c \eta}{2 \alpha_d}.
\end{equation}
Notably, the obtained attenuation length depends on the square of the
structure period, $l'' \propto d^2$, similarly to its behavior in electrodynamic
spoof-plasmonic structures~\cite{rusinaTheory2010}.

\subsection{Transmission spectra}

Overall transport properties of a finite structures inside the interval
$\left[ X_A, X_B \right] $ are described by the reflection and transmission
spectra obtained by solving the equations of motion assuming the scattering
boundary conditions. In terms of transfer matrices, the solution has the
form $\mat{\Psi}(X_B) = \mathcal{T}(X_B, X_A) \mat{\Psi}(X_A)$. In turn, the states at the
terminating points are written in the basis of independent solutions as 
$\mat{\Psi}^{(w)}(X_A) = \mathbf{w}_+ +
r_{BA}\mathbf{w}_-$ and $\mat{\Psi}^{(w)}(X_B) = t_{BA}\mathbf{w}_+$, where
$\mathbf{w}_+ = \mqty(1, & 0)^T$ and $\mathbf{w}_- = \mqty(0, & 1)^T$, and $r_{BA}$
and $t_{BA}$ are the reflection and transmission coefficients,
respectively:
\begin{equation}\label{eq:tran_refl_coeffs}
  \begin{split}
    t_{AB} & = \frac{\det \mathcal{T}(X_B, X_A)}{
             \left( \mathbf{w}_-, \mathcal{T}(X_B, X_A) \mathbf{w}_-\right)}, \\ 
   r_{AB} & = - \frac{\left( \mathbf{w}_-, \mathcal{T}(X_B, X_A)
            \mathbf{w}_+\right)}{
            \left( \mathbf{w}_-, \mathcal{T}(X_B, X_A) \mathbf{w}_-\right)}.
  \end{split}
\end{equation}
Here, we have defined the inner product as $(\mathbf{a}, \mathbf{b}) :=
\sum_{j} a^*_j b_j$ and kept $\det \mathcal{T}(X_B, X_A) = 1$.

\subsection{External excitations}

The external source can be incorporated into the system dynamics by adding
terms $-F(x)$ and $-F_n$ to the right-hand-side of the first and the second
equations of Eqs.~\eqref{eq:string_eigen}, respectively. We consider the
effect of external sources under ``physical'' boundary conditions assuming
that local resonances and sources are strictly inside of the interval
$\left[ X_A, X_B\right]$, while at the boundary of the interval the
excitations satisfy certain boundary conditions. In the present work, we
limit our attention only to outgoing boundary conditions so that
$\boldsymbol{\Psi}^{(w)}\left( X_{A} \right) \propto \mathbf{w}_-$ and
$\boldsymbol{\Psi}^{(w)}\left( X_{B} \right) \propto \mathbf{w}_+$.

The general solution of the equations of motion with the source at points
outside of the resonances can be written in the basis-independent form as
\begin{equation}\label{eq:EE-fundamental}
  \mat{\Psi}\left( x_2 \right) =
  \mathcal{T}\left( x_2, x_1 \right) \mat{\Psi}\left( x_2 \right) +
  \int_{x_1}^{x_2} \dd{s} \mathcal{T}\left( x_2, s \right) F(s) \mathbf{V}(s),
\end{equation}
where $\mathbf{V}(s)$ is the vector accounting for the source within the
transfer matrix formalism. In the Cauchy basis and the basis of independent
solutions, the source vector is 
\begin{equation}\label{eq:exc-vectors}
  \begin{split}
   \mathbf{V}^{(C)}(s) & = \mqty(0 \\ 1), \\ 
    \mathbf{V}^{(w)}(s)  &  = \frac{1}{2\rmi k}
                           \mqty(\rme^{- \rmi k s} \\ -\rme^{\rmi k s}).
  \end{split}
\end{equation}

We will assume for simplicity that the source is applied outside of
resonances ($F_n \equiv 0$). It should be noted that this assumption does not
loose generality, if under the integral in Eq.~\eqref{eq:EE-fundamental}
for $x_2 > x_1$, one adopts the convention
$\widehat{T}\left( x_2, s \right) = \widehat{T}\left( x_2, s + 0\right)$
and correctly defines the respective $F(s)$. Namely, let the $n$-th
resonance be excited with the source of amplitude $F_n$. Then, its
contribution to $\mat{\Psi}^{(C)}(x_2)$ in~\eqref{eq:EE-fundamental} written
in the Cauchy basis has the form
\begin{equation}\label{eq:mech-ext-nth-res}
  \gamma(\omega)F_n \widehat{T}^{(C)} \left( x_2, x^{(n)}+0 \right) \mathbf{V}^{(C)}(s),
\end{equation}
where transfer matrix $\widehat{T}^{(C)}\left( x_2, x^{(n)}+0 \right) $
does not include the transfer matrix across the $n$-th resonance itself.

In what follows we will be specifically interested in the situation when
the field is detected at the terminal points (output ports) of the
structure. For such situations, the effect of the external excitation can
be expressed by a single source vector. It is straightforward to check that
choosing a point $\overline{X}$ inside the structure,
$X_A < \overline{X} < X_B$, one has
\begin{equation}\label{eq:EE-representative}
\mat{\Psi}\left( X_B \right) =
  \mathcal{T}\left( X_B, X_A \right) \mat{\Psi}\left( X_A \right) +
  \mathcal{T}\left( X_B, \overline{X} \right)
  \overline{\mathbf{V}}\left( \overline{X} \right) ,  
\end{equation}
where we introduced a representing source vector
\begin{equation}\label{eq:EE-representative-def}
 \overline{\mathbf{V}}\left( \overline{X} \right)  = 
\int_{X_A}^{X_B} \dd{s} \mathcal{T}\left( \overline{X}, s \right) F(s) \mathbf{V}(s).
\end{equation}

To illustrate the difference between the free (without the local
resonances) and spoof-plasmonic systems, we consider the effect of canceled
output of two point sources separated by a subwavelength distance. The
outgoing boundary conditions ensure that this effect takes place when
$\overline{\mathbf{V}} = 0$. 
Let the sources be applied at points $s_1 = -d_1$ and $s_2 = d_2$, with
positive $d_1$ and $d_{2}$, so that
$F(s) = f_1 \delta(s - s_1) + f_2 \delta(s - s_2)$. Then, writing
Eq.~\eqref{eq:EE-representative-def} in the basis of independent solutions,
we obtain for the free system
\begin{equation}\label{eq:EE-cancel_free_rep}
  \overline{\mathbf{V}}_{\text{free}} = f_1 \mathbf{V}(s_1)
  + f_2 \mathbf{V}(s_2).
\end{equation}
The condition of existence of nontrivial solutions to
$\overline{\mathbf{V}}_{\text{free}} = 0$ is the easiest to obtain by
writing the source representing vector in the basis of independent
solutions: $\sin(k(d_1 + d_2)) = 0$. Thus, for close sources, with
$k(d_1 + d_2) < \pi$, this condition cannot be met.

The presence of local resonances in the spoof-plasmonic channel changes the
situation drastically. It should noted that if the source locations are not
separated by a local resonance, one can choose $\overline{X} > s_2$ in such
a way that the interval $[s_1, \overline{X}]$ does not contain local
resonances. In this case, the source representing vector is 
determined by the same Eq.~\eqref{eq:EE-cancel_free_rep} as for the free
system, leading to the no-cancellation outcome. 

Let the sources be placed in the same way as for the free system with the
local resonance between them, $x^{(0)} = 0$. Then, we have
\begin{equation}\label{eq:EE-cancel_SP_rep}
  \overline{\mathbf{V}}_{\text{SP}} =
  f_1 \mathcal{T}_0 \mathbf{V}(s_1)
  + f_2 \mathbf{V}(s_2),
\end{equation}
where we have introduced $\mathcal{T}_0 = \mathcal{T}\left( x^{(0)} \right)$.

The condition for the existence of a non-trivial solution to
$\overline{\mathbf{V}}^{(w)}_{\text{SP}} = 0$ is found to be
\begin{equation}\label{eq:EE-cancel_cond_gen}
  \sin \left( k(d_1 + d_2) \right) + \frac{\gamma(\omega)}{2k} \sin (kd_1) \sin (kd_2) = 0,
\end{equation}
which can be satisfied even for subwavelength separation of the sources.

\begin{figure}[tb]
  \centering
  \includegraphics[width=3in]{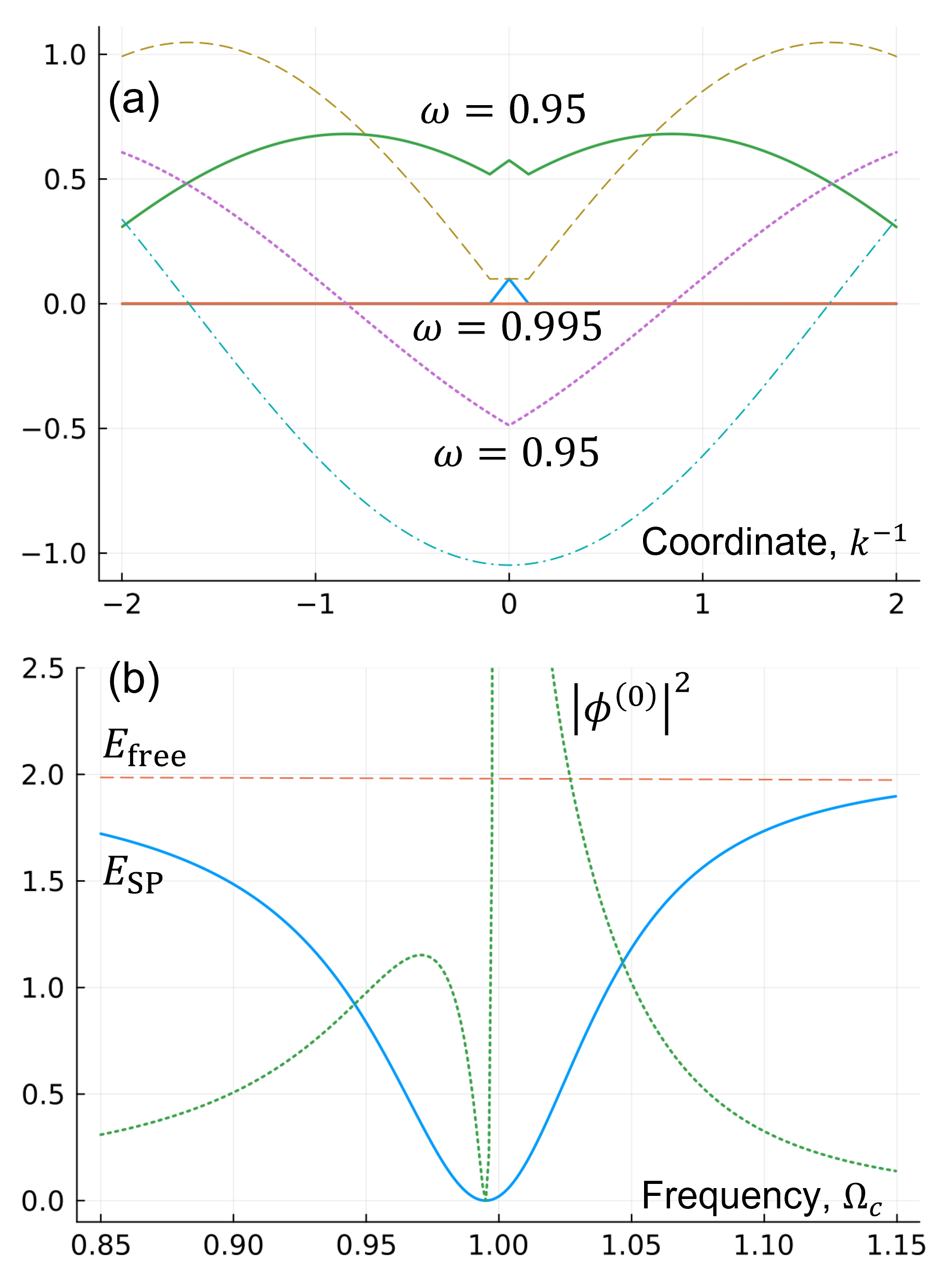}
  \caption{The two-source cancellation in spoof-plasmonic channels. (a) The
    response of the free system (dashed and dashed dotted lines show,
    $\mathrm{Re}[\psi(x)]$, and $\mathrm{Im}[\psi(x)]$, respectively) at
    $\omega/\Omega_c = 0.95$, and spoof-plasmonic systems at
    $\omega/\Omega_c = 0.95$ (the cancellation condition is not met) and
    $\omega/\Omega_c = 0.995$ (the condition is met). Solid and dotted lines show
    $\mathrm{Re}[\psi(x)]$ and $\mathrm{Im}[\psi(x)]$, respectively. (b) The
    frequency dependence of the outgoing signal
    $E_j(\omega) = \abs{\psi(X_A)}^2 + \abs{\psi(X_B)}^2$ with
    $j = \text{free}$ (dashed line) and $j = \text{SP}$ (solid line) for
    the free and spoof-plasmonic systems, respectively. The dotted line
    shows the magnitude of the local resonance excitation
    $\abs{\phi^{(0)}(\omega)}^2$.}
  \label{fig:cancellation}
\end{figure}

We consider in more detail the case of symmetric placement of sources
around the resonance: $d_1 = d_2 = d$. In this case, the cancellation
condition has the form $B(\omega; d) = 0$ and 
corresponds to the edge of the spoof-plasmon fundamental band in the
structure with period $d$. When the condition is met, the nontrivial
solution to Eq.~\eqref{eq:EE-cancel_SP_rep} is $f_1 = f_2 \ne 0$.
Figure~\ref{fig:cancellation} compares the response of the system when the
cancellation condition is met and the system is tuned away from the
cancellation frequency.

It is worth emphasizing that the cancellation occurs at the edge of the
fundamental spoof-plasmonic band [Eq.~\eqref{eq:mech-band-edge}]. This is
an example of the emergence of spoof-plasmonic features that are commonly
associated with periodic systems in short systems containing only few local
resonances~\cite{joySpoof2017}.


\section{Near-field-assisted information channel capacity}
\label{sec:info-cap}

A source applied inside the structure creates a signal at both terminating
points of the structure implying the information flow from the source to
the output ports. An important feature of spoof-plasmonics systems is
revealed when one considers the source distributed over a region of a
subwavelength size. In this case, the information flow can be regarded as
carried by two simple channels activated by the long-scale and the
short-scale features of the source distribution. In the free system, the
short-scale channel remains only weakly activated owing to the small size
of the excitation region. However, in the spoof-plasmonic systems, in the
frequency region where the plasmonic features become prominent, the
contribution of the short-scale channel can significantly increase.

The capacity of the channel between the source $S$ and the output ports
$AB$ is defined as $\max_{P_S} I(AB : S)$, where $I(AB : S)$ is the mutual
information of the source and the output ports, and the maximum is taken
over the distribution functions describing the source, $P_S$. This
definition can be simplified by taking into account that, by virtue of the
outgoing boundary conditions, the output state is given in the basis of
independent solutions by $\mat{\Psi}^{(w)}(X_A) = c_- \mathbf{w}_-$ and
$\mat{\Psi}^{(w)}(X_B) = c_- \mathbf{w}_+$. It is convenient to represent the
state in a vector form
$\mathbf{c} = \left( c_+, c_- \right)^T \in \mathbb{C}^2$. Consequently, the
same situation as above, with two point sources, is sufficient to consider the
channel capacity without loss of generality as long as vectors
$\mathbf{V}^{(w)}(s_1)$ and $\mathbf{V}^{(w)}(s_2)$ are linearly
independent. Introducing the vector notation for the source amplitudes
$\mathbf{f} = \left( f_1, f_2 \right)^T \in \mathbb{C}^2$, we have a relation
between the output signal and the source
$\mathbf{c} = \widehat{G} \mathbf{f}$. The matrix elements of $\widehat{G}$
(the transfer function) are found using Eqs.~\eqref{eq:EE-representative}
and~\eqref{eq:EE-representative-def}
\begin{equation}\label{eq:IC-GG-form}
  \begin{split}
    G_{1j} & = t_{AB}
             \left( \mathbf{w}_+, \mathcal{T}(X_A, s_j) \mathbf{V}^{(w)}(s_j) \right), \\ 
    G_{2j} & = - t_{AB}
             \left( \mathbf{w}_-, \mathcal{T}(X_B, s_j) \mathbf{V}^{(w)}(s_j) \right),
  \end{split}
\end{equation}
where $j = 1, 2$. While deriving~\eqref{eq:IC-GG-form}, we have taken into
account that $\mathcal{T}(X_A, X_B) = \mathcal{T}^{-1}(X_B, X_A)$ and the relation between the
matrix elements of $2\times2$ matrices: $T_{22} = \left( T^{-1} \right)_{11}$.


Such obtained relation between the output and the sources allows one to use
the standard approach for multivariate Gaussian channels and write down the
channel capacity as (see, e.g. Ref~\onlinecite{foschiniLimits1998,
  telatarCapacity1999})
\begin{equation}\label{eq:IC-full-channel-capacity}
  C_{\mathrm{SP}} = \max_{\widehat{K}_f}
  \frac{1}{2}
  \log \left[ \det \left( 1 +
      \frac{S}{\nu} \widehat{G} \widehat{K}_f \widehat{G}^\dagger \right) \right],
\end{equation}
where $S/\nu$ is the signal-to-noise ratio,
$\widehat{K}_f = \expval{\mathbf{f} \times \mathbf{f}^\dagger}$ is the (normalized)
source covariance matrix, and the maximum is taken over all admissible
covariance matrices. 

Individual subchannels contributing to $C_{\mathrm{SP}}$
in~\eqref{eq:IC-full-channel-capacity} can be identified as the basis, in
which transfer function $\widehat{G}$ is diagonal, so that $g^{(j)}$, the
eigenvalues of $\widehat{G}$, have the meaning of the subchannel scalar
transfer functions. For example, assuming that the input does not mix the
channels, Eq.~\eqref{eq:IC-full-channel-capacity} turns into
$C_{\mathrm{SP}} = \sum_{j} C_{\mathrm{SP}}^{(j)}$, where $j$ runs over the
channels, and $ C_{\mathrm{SP}}^{(j)}$ is the capacity of the $j$-th
subchannel:
\begin{equation}\label{eq:IC-subchannel-capacity}
 C_{\mathrm{SP}}^{(j)} = 
  \frac{1}{2}
  \log \left( 1 + \frac{\expval{\abs{f^{(j)}}^2}}{\nu} \abs{g^{(j)}}^2 \right).
\end{equation}

With this perspective in mind and to avoid the application-specific
questions regarding the choice of the source covariance matrix, we
limit our consideration to the transfer functions of individual channels.
To make apparent the emergence of channels associated with different scales
of the source spatial variation, we consider the case, when the systems and
the source placement has the inversion symmetry. In other words, we
consider sources placed at $s_{1,2} = \pm d$ in a system with local
resonances at $x^{(n)} = p n$ with $p > d$ and $-N \leq n \leq N$.

In this case, the transfer function $\widehat{G}$ preserves the symmetry
(symmetric and antisymmetric distributions) of the source and is
diagonalized by $\widehat{M} = \mqty(1 & 1 \\ 1 & -1)/\sqrt{2}$
\begin{equation}\label{eq:IC-G-diagonal}
  \widetilde{G} = \widehat{M} \widehat{G} \widehat{M}^{-1}
  = \mqty(g^{(+)} & 0 \\ 0 & g^{(-)}),
\end{equation}
where the eigenvalues $g^{(\pm)}$ have the meaning of the scalar transfer
functions corresponding to the channels activated by the symmetric and
antisymmetric sources. From the perspective of the source spatial
distribution, $g^{(+)}$ transfers the signal corresponding to the spatially
averaged source, while $g^{(-)}$ carries the signal determined by the
short-scale features of the source distribution.

\begin{figure}[tb]
  \centering
  \includegraphics[width=3in]{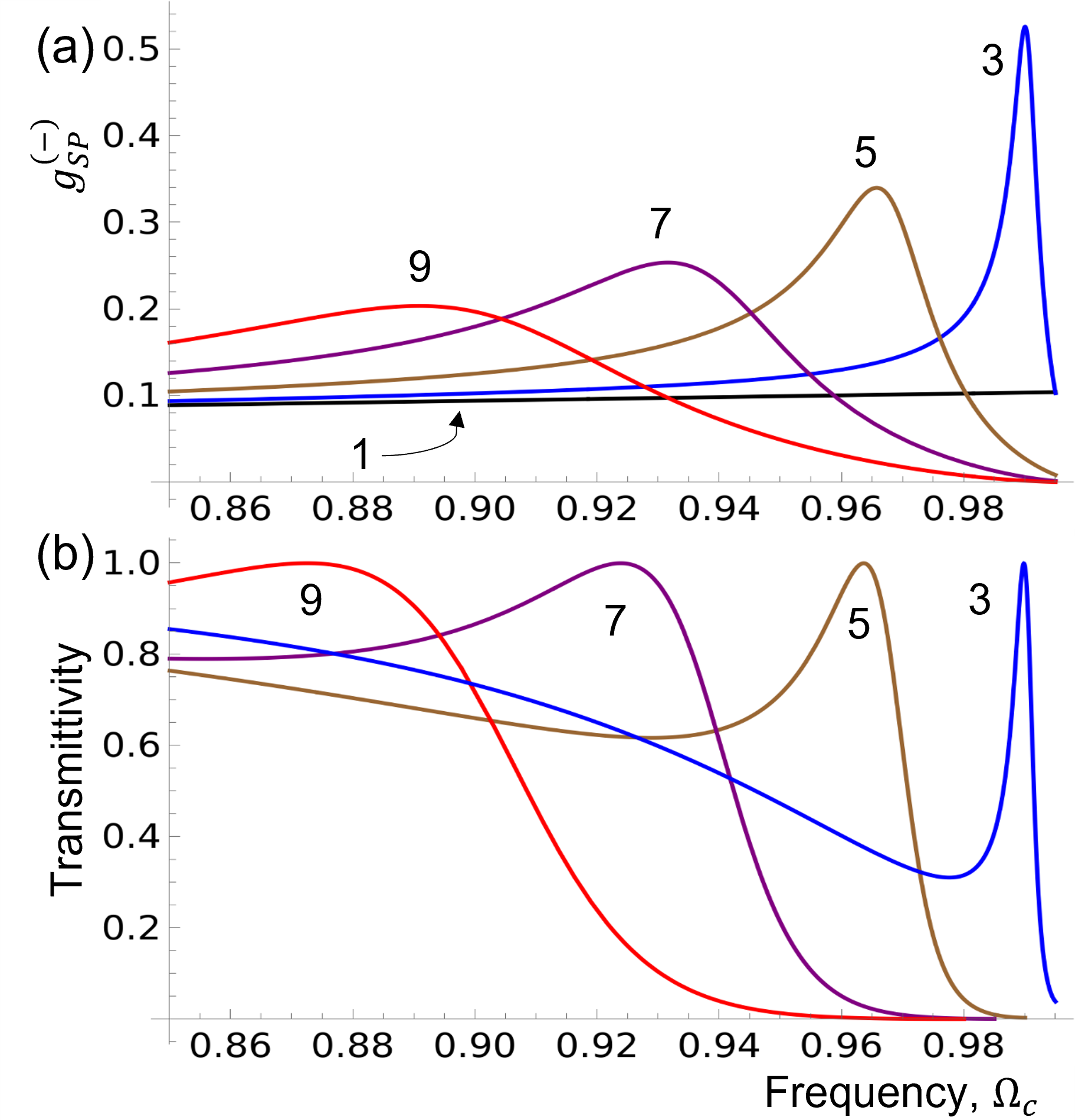}
  \caption{(a) The frequency dependence of the scalar transfer function
    $g^{(-)}_{\mathrm{SP}}(\omega)$ carrying the signal induced by the details
    of the source spatial variation for structures with
    $2N + 1 = 3, 5, 7, 9$ local resonances. The separation between the
    sources and the local resonances is chosen $0.1$~$1/k_c$. The line
    corresponding to only one local resonance coincides with the transfer
    function of the free system. (b) The transmission spectrum,
    $\abs{t_{AB}(\omega)}^2$, for the same structures.}
  \label{fig:short-scale-channel}
\end{figure}

This distinction becomes apparent for the free system, for which one has
$g^{(\pm)} = \left( \mathbf{w}_{\pm}, \mathbf{V}^{(w)}(s_1) \pm
  \mathbf{V}^{(w)}(s_2) \right) $, so that
$g^{(+)}_{\mathrm{free}}(d) =  \cos(kd)/\rmi k$ and
$g^{(-)}_{\mathrm{free}}(d) = \sin(kd)/k$. For sources confined to a
subwavelength interval, one has $g^{(-)}_{\mathrm{free}}(d) \ll 1$, and the
short-scale channel remains under-activated.

In the spoof-plasmonic system, the capacity of the short-scale channel is
modified, compared to the free system, by the resonances. Introducing
$\mathcal{T}(X_B, x^{(0)}) \equiv \mathcal{T}(X_B, x^{(0)} + 0)\mathcal{T}_0^{1/2}$, we obtain our main result
\begin{equation}\label{eq:IC-gm-sp}
  g^{(-)}_{\mathrm{SP}} = g^{(-)}_{\mathrm{free}}(d) t_{AB}
  \left( \mathbf{w}_-, \mathcal{T}^{(w)}\left( X_B, x^{(0)} \right)  \mqty(1 \\ 1) \right).
\end{equation}

It is constructive to consider the case when the structure, similarly to
the consideration of the cancellation effect, contains only one local
resonance. In this case, one has $t_{AB} = (1 + \rmi \gamma/2k)^{-1}$, and, as a
result, the transfer function coincides with that for the free system. This
is utilized in Fig.~\ref{fig:short-scale-channel} illustrating that the
overall frequency dependence of $g^{(-)}_{SP}$ is mainly determined by the
transmission spectrum of the structure.

To show the enhancement more explicitly, we use the fact that the local
resonance at $x^{(0)}$ does not affect the contribution of the
antisymmetric component and consider the simplest structure where the
enhancement takes place: with resonances at $x^{(\pm 1)} = \pm p$ and sources
at $s_{1,2} = \mp d$. For this case, we obtain
\begin{equation}\label{eq:IC-simple-enhancement}
g^{(-)}_{\mathrm{SP}}(d) = g^{(-)}_{\mathrm{free}}(d) \frac{\rme^{-\rmi k
  p}}{B_2(\omega; p) + \rmi \sin(kp)},
\end{equation}
where $B_2(\omega; p) = \cos(kp) + (\gamma/k) \sin(kp)$ has the meaning of the
dispersion equation of spoof plasmons with the modified coupling constant
$\eta \to 2\eta$. Equation~\eqref{eq:IC-simple-enhancement} shows that
$g^{(-)}_{\mathrm{free}}$ is modified by a resonance centered at the edge
of the fundamental band of the modified and with the width
$2 k_c \sin^2(k_c p)/\eta\cos(k_c p)$. We find that the maximal value of the
enhancement factor is
$\abs{g^{(-)}_{\mathrm{SP}} / g^{(-)}_{\mathrm{free}}} = 1/\sin(k_c p)$.
For example, using $p = 0.1 k_c^{-1}$, as in
Fig.~\ref{fig:short-scale-channel}, we obtain about five-fold enhancement.

The dependence of the transfer function of the long-scale channel,
$g^{(+)}_{SP}$, on the separation of the sources reflects the cancellation
effect discussed above
\begin{equation}\label{eq:IC-gp-sp}
    g^{(+)}_{\mathrm{SP}} = \frac{B(\omega; d)}{\rmi k} 
  t_{AB} \left( \mathbf{w}_-, \mathcal{T}^{(w)}\left( X_B, x^{(0)} \right)
      \mqty(1 \\ -1)\right).
\end{equation}

\begin{figure}[tb]
  \centering
  \includegraphics[width=3in]{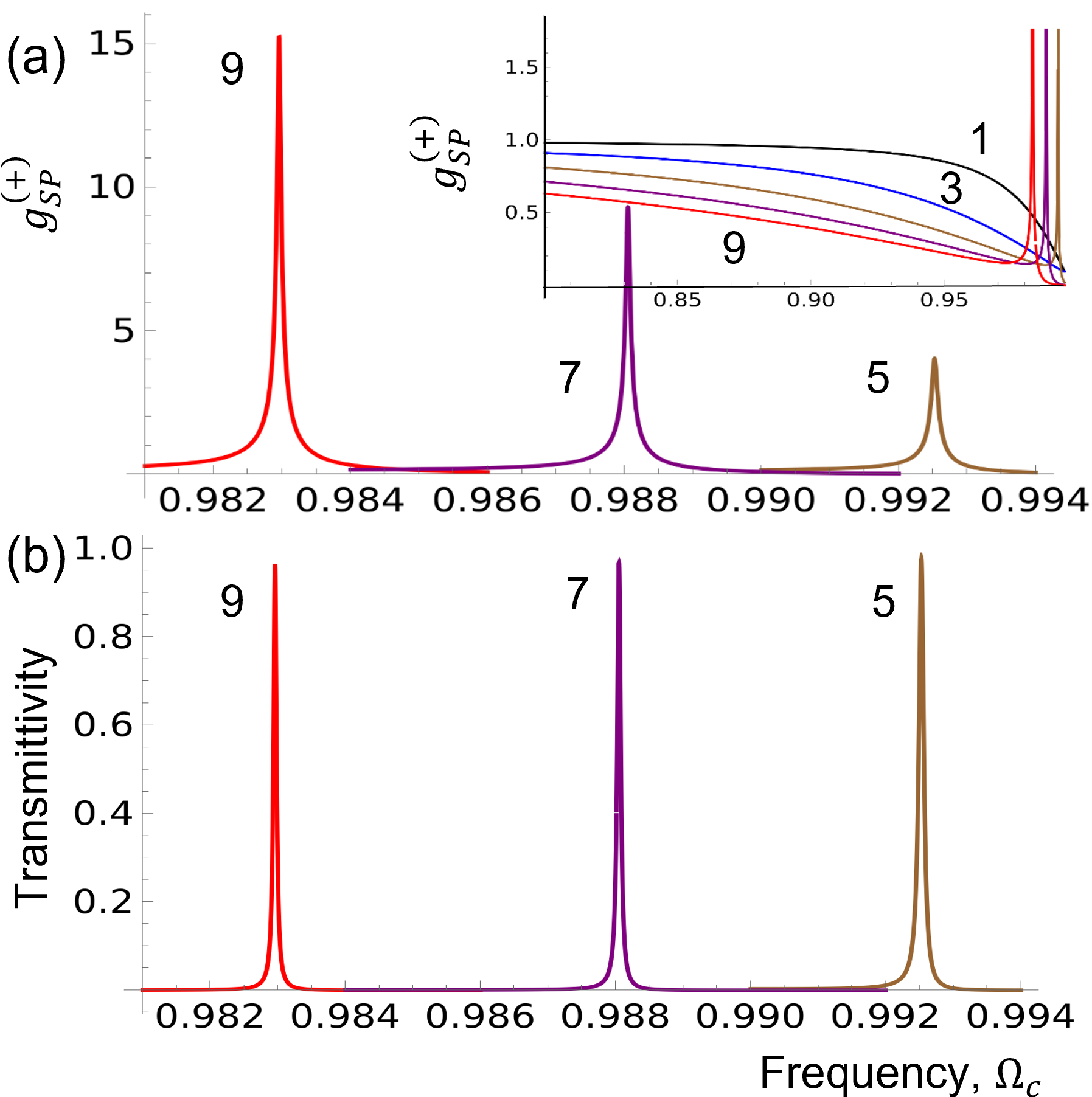}
  \caption{(a) The fine details of the frequency dependence of the scalar
    transfer function $g^{(+)}_{\mathrm{SP}}(\omega)$ near the resonances for
    structures with $2N + 1 = 5, 7, 9$ local resonances. The inset shows
    $g^{(+)}_{\mathrm{SP}}(\omega)$ in a wide frequency interval. (b) The
    transmission resonances, $\abs{t_{AB}(\omega)}^2$, responsible for
    resonances in the transfer function.}
  \label{fig:long-scale-channel}
\end{figure}

A curious feature of $g^{(+)}_{\mathrm{SP}}(\omega)$ is the emergence of
narrow resonances near the edge of the fundamental spoof-plasmonic band.
Similarly to $g^{(+)}_{\mathrm{SP}}(\omega)$, these resonances are caused by
transmission resonances, as illustrated by
Fig.~\ref{fig:long-scale-channel}. It must be noted, however, that due to
the proximity of these resonances to the edge of the fundamental band, they
can be expected to be strongly affected by the decay. As a result, the
significance of these resonances in particular physical implementations of
spoof-plasmonic channels may be significantly reduced and should be studied
on the case-by-case basis.

We conclude by noticing that the inversion symmetry simplifies finding the
explicit form of the spoof plasmonic modes of the structure. The role of
the symmetrical placement of the sources is secondary as any source can be
decomposed as a superposition of contributions with definite symmetry
(projections on the respective irreducible representations). Indeed, using
the notion of the transfer matrix from the center of the structure,
$\mathcal{T}(X_B, x^{(0)})$, we can choose $\overline{X} = x^{(0)}$ and define the
source representing vector at the center of the structure
$\overline{\mathbf{V}}_0 = \overline{\mathbf{V}}\left( x^{(0)} \right)$, with
\begin{equation}\label{eq:IC-rep-center}
  \overline{\mathbf{V}}^{(w)}\left( x^{(0)} \right) =
  f_1 \mathcal{T}_0^{1 / 2} \mathbf{V}(s_1)
  + f_2 \mathcal{T}_0^{-1 / 2} \mathbf{V}(s_2),
\end{equation}
so that the last term in Eq.~\eqref{eq:EE-representative} has the form
$\mathcal{T}(X_B, x^{(0)})\overline{\mathbf{V}}_0$. Thus, we obtain for
$s_1 = x^{(0)}-d_1$ and $s_2 = x^{(0)} + d_2$
\begin{equation}\label{eq:IC-rep-center-explicit}
\begin{split}
  \overline{\mathbf{V}}^{(w)}_0 =
  & 
    \frac{1}{2\rmi k} \left[ f_1 B(\omega; d_1) + f_2 B(\omega; d_2)\right]
    \mqty(1 \\ -1)
  \\ 
  & \frac{1}{2 k} \left[ f_1 \sin(k d_1) - f_2 \sin(k d_2)\right]
    \mqty(1 \\ 1).
\end{split}  
\end{equation}
Here, the first and the second terms in the right-hand side represent
symmetric and antisymmetric components, respectively.
Representation~\eqref{eq:IC-rep-center-explicit} explicitly expresses the
fact that the local resonance placed between the sources does not affect
the antisymmetric component, as has been used above for evaluating the
enhancement factor.

\section{Conclusion}

We considered a simplified 1D scalar model of spoof-plasmonic channels.
Mechanically, the model can be represented as a string with attached
harmonic oscillators playing the role of local leaky resonances.
We show that the model reproduces main spectral feature of spoof-plasmonic
channels, including such subtle effects as a the red-shift of the edge of
the fundamental band from the resonance frequency and the dependence of the
attenuation length on the square of the structure period.

Owing to its simplicity, the model allows investigating features that are
to challenging to establish within the frameworks of first-principle
descriptions. We use this opportunity to investigate manifestations of
plasmonic-like features in structures with only few local resonances.

The main result is the enhanced sensitivity of the signal induced in
the structure by the source on the spatial variation of the source at the
subwavelength scale. The sensitivity is characterized from the perspective
of the informational capacity of the channel between the source and the
output ports (terminating ends of the structure). It is shown that while
this channel comprises two subchannels, the activation of one of
subchannels depends on the size of the region where the source is applied.
As a result, if the source region is of the subwavelength size, the
subchannel is characterized by a small channel capacity in conventional
systems. However, spoof-plasmonic structures demonstrate enhancement of the
short-scale subchannel in the frequency region where the spoof-plasmonic
effects are prominent.

From a more general perspective, this effect demonstrates that the flow of
information carried by waves can be significantly impacted by strong
scattering resonances even if only a few of such resonances are present.

\section*{Acknowledgment}

The work has been supported by the US National Science Foundation (NSF)
under Grant No. 1909937.

\appendix

\section{Bases in the transfer matrix formalism}
\label{sec:tm-approach}

The transfer matrix $\mathcal{T}(x_2, x_1)$ relates state vectors at two points
\begin{equation}\label{eq:tm-tm_cauchy}
\mat{\Psi}(x_2) = \mathcal{T}(x_2, x_1) \mat{\Psi}(x_1).
\end{equation}
The main actively employs the fact that the state vector can be specified
in different ways to better accommodate the specifics of a particular
problem. The explicit form of the transfer matrix depends on the choice of
the representation of the local state, which can be regarded as choosing
the basis for the state vector. Here, we provide the relation between the
transfer matrices given in two bases used in the main text.

The first basis utilizes the observation that a solution of the second
order ODE with respect to $\psi(x)$ is specified
by providing the pair $\psi(x_1)$ and $\psi'(x_1)$ at some point $x_1$.
This leads to the state vector
\begin{equation}\label{eq:tm-cachy_psi}
  \mat{\Psi}^{(C)}(x) = \mqty( \psi(x) \\ \psi'(x)).
\end{equation}
Within the remaining parts of the interval, $\psi(x)$, and hence $\mat{\Psi}(x)$,
is found by solving the governing ODE as the Cauchy problem treating
$\mat{\Psi}(x_1)$ as the initial condition:
$\mat{\Psi}^{(C)}(x_2) = \mathcal{T}^{(C)}(x_2, x_1) \mat{\Psi}^{(C)}(x_1)$.
Because of this, we will refer to $\mat{\Psi}^{(C)}(x)$ and $\mathcal{T}^{(C)}(x_2, x_1)$
as written in the Cauchy basis.

An alternative approach to specifying the state vector uses a pair of
linearly independent functions, $h_{1,2}(x)$:
\begin{equation}\label{eq:tm-funct_psi}
 \mat{\Psi}^{(h)}(x) = \mqty( c_1(x) \\ c_2(x)),
\end{equation}
where the superscript $h$ signifies the functions pair, and the amplitudes
$c_{1,2}(x)$ are found as the solution of the system of equations
\begin{equation}\label{eq:tm-general_basis_definition}
\begin{split}
\psi(x) = c_1(x) h_1(x) + c_2(x) h_2(x), \\
\psi'(x) = c_1(x) h'_1(x) + c_2(x) h'_2(x).
\end{split}
\end{equation}
The existence of solutions is warranted by the linear independence of vectors
$(h_1(x), h_1'(x))^T$ and $(h_2(x), h_2'(x))^T$, that is at points where
the Wronsky matrix
\begin{equation}\label{eq:tm-Wronsky_matrix_definition}
\widehat{W}_h(x) = \begin{pmatrix}    h_{1}(x) & h_{2}(x) \\
h'_{1}(x) & h'_{2}(x)  \end{pmatrix}.
\end{equation}
is nonsingular.

A particularly convenient choice is the pair
of independent solutions of the governing ODE: $h_1 = w_+(x)$, $h_2 =
w_-(x)$. For example, for a free wavesystem at the frequency corresponding
to wavenumber $k$, it is convenient to choose $w_{\pm}(x) = \rme^{\pm \rmi
  kx}$, in which case the Wronsky matrix takes the form
\begin{equation}\label{eq:tm-Wronsky-matrix-solutions}
  \widehat{W}_w(x) = \mqty( \rme^{\rmi kx} & \rme^{-\rmi kx} \\
  \rmi k \rme^{\rmi kx} & -\rmi k \rme^{-\rmi kx}).
\end{equation}
We will say that the state vector is written in the basis of independent
solutions. In this basis the transfer matrix is trivial
$T^{(w)}(x_2, x_1) = \widehat{1}$.

Equations~\eqref{eq:tm-general_basis_definition} provide the relation
between state vectors in different bases:
\begin{equation}\label{eq:relation_state_vectors}
  \mat{\Psi}^{(C)}(x) = \widehat{W}_h(x) \mat{\Psi}^{(h)}(x),
\end{equation}
which leads to
\begin{equation}\label{eq:relation_field_functions}
\mathcal{T}^{(h)}(x_2, x_1) = \widehat{W}_h^{-1}(x_2)\mathcal{T}^{(C)}(x_2,x_1) \widehat{W}_h(x_1).
\end{equation}
We note that the relation between transfer matrices in
different bases is not a similarity transformation.


\section{Equivalence of different forms of the dispersion equation}
\label{sec:app-equivalence}

To show the equivalence of
\begin{equation}\label{eq:app-long_disp}
  \cos(\beta d) = \cos(kd) + \frac{\gamma}{2k} \sin(kd).
\end{equation}
and
\begin{equation}\label{eq:app-bloch_disp}
 1 = \frac{\gamma}{d} \sum_{m} \frac{1}{k^2 - \beta_m^2},
\end{equation}
with $\beta_m = \beta + 2 \pi \frac{m}{d}$, we, first, rewrite
Eq.~\eqref{eq:app-bloch_disp} as
\begin{equation}\label{eq:mech_tra_tm-bragg-1}
\begin{split}
 \frac{2k}{\gamma} &  = \frac{1}{\varphi_+} + \frac{1}{\varphi_-} +\\ 
  & \sum_{m>0} \left( \frac{2 \varphi_+}{\varphi_+^2 - (2\pi m)^2}
  + \frac{2 \varphi_-}{\varphi_-^2 - (2\pi m)^2}\right),
\end{split}
\end{equation}
where $\varphi_\pm = d(k \pm \beta)$. Next, we observe that
\begin{equation}\label{eq:mech_tra_tm-bragg-diff}
  \frac{2\varphi}{\varphi^2 - (2 \pi m)^2} = \frac{\partial}{\partial\varphi}
      \ln \left( 1 - \left( \frac{\varphi}{2\pi m} \right)^2 \right).
\end{equation}
Thus, we have for the sum in the r.h.s. of~\eqref{eq:mech_tra_tm-bragg-1}
\begin{equation}\label{eq:mech_tra-tm-bragg-sums}
\begin{split}
  \sum_{m > 0} =   &  \frac{\partial}{\partial\varphi_+}
      \ln \left[ \prod_{m=1}^\infty \left( 1 - \left( \frac{\varphi_+}{2\pi m} \right)^2
        \right) \right]  + \\ 
  &   \frac{\partial}{\partial\varphi_-}
      \ln \left[ \prod_{m=1}^\infty \left( 1 - \left( \frac{\varphi_-}{2\pi m} \right)^2 \right) \right]
\end{split}  
\end{equation}
Using the Euler infinite product representation,
\begin{equation}\label{eq:mech_Euler}
  \frac{\sin(\pi z)}{\pi z} = \prod_{m > 0} \left( 1 - \left( \frac{z}{m} \right)^2 \right),
\end{equation}
we obtain
\begin{equation}\label{eq:mech_tra_tm-bragg-2}
\begin{split}
  \frac{2k}{\gamma} &
     = \frac{1}{\varphi_+} + \frac{1}{\varphi_-} + \\ 
  &   \frac{\partial}{\partial\varphi_+} \ln \left[ \frac{\sin(\varphi_+/2)}{\varphi_+/2} \right]
  + \frac{\partial}{\partial\varphi_-} \ln \left[ \frac{\sin(\varphi_-/2)}{\varphi_-/2} \right],
\end{split}  
\end{equation}
which simplifies to 
\begin{equation}\label{eq:mech_tra_tm-bragg-2-1}
  \frac{2k}{\gamma} = 
 \frac{\cos(\varphi_+/2)}{2\sin(\varphi_+/2)}
  + \frac{\cos(\varphi_-/2)}{2\sin(\varphi_-/2)}.
\end{equation}
This expression is straightforward to reshape to
form~\eqref{eq:app-long_disp}.


%

\end{document}